\documentclass[%
 aip,
cp,  
 amsmath,amssymb,
reprint,%
]{revtex4-2}
\usepackage{graphicx}
\usepackage{dcolumn}
\usepackage{bm}

\usepackage[utf8]{inputenc}
\usepackage[T1]{fontenc}
\usepackage{mathptmx} 

\begin{document}

\title{Recent Results on Charmonia- and Bottomonia-like Particles\\ at Belle}

\author{Pin-Chun Chou on behalf of the Belle Collaboration}
\affiliation{
  Department of Physics, National Taiwan University, Taipei 10617, Taiwan
}
\date{\today} 

\begin{abstract}
The large data sample accumulated by the Belle experiment at KEKB asymmetric energy $e^+ e^-$ collider provides opportunities to study charmonia (bottomonia) and charmonium-like (bottomonium-like) exotic particles. In this review, we report recent results on these topics from Belle, including searches for $B\to h_c K$, $B\to Y(4260)K$, $B\to X(3872/3915)  (\to \chi_{c1} \pi^0 )  K$, $B^0 \to X(3872) \gamma$, $e^+e^- \to \gamma \chi_{cJ}$ and a new measurement of the $e^+ e^-\to \Upsilon(nS) \pi^+ \pi^-  (n=1,2,3)$ cross sections at energies from 10.52 to 11.02 GeV.
\end{abstract}

\maketitle

\section{\label{sec:intro}Introduction}
Since the discovery of the charmonium-like exotic $X(3872)$ state at the Belle experiment in 2003~\cite{ref:PRL2003}, many new conventional quarkonium states and more than a dozen of exotic quarkonium-like states were discovered. 
The charmonium and charmonium-like states have been intensively studied via the $B$ decays as well as the initial state radiation (ISR) process at Belle, and the bottomonium spectroscopy was investigated by the state transitions. 
In this review, the study for some charmonium and charmonium-like states and a new measurement of the cross sections for $e^+ e^-\to \Upsilon(nS) \pi^+ \pi^-  (n=1,2,3)$ at Belle are reported.
Charge-conjugated modes are implied throughout this review. 

\section{Search for $B \to Y(4260) K$\cite{ref:Y4260}}
The branching fraction product $\mathcal{B}(B^+ \to Y(4260) K^+) \times \mathcal{B}(Y(4260) \to J/\psi \pi^+ \pi^-)$ is predicted to be in the range $3.0\times10^{-8} - 1.8 \times 10^{-6}$ by a QCD sum-rule model which assumes that the $Y(4260)$ is a mixture of charmonium and tetraquark states~\cite{ref:PLB74783}.
In 2006, the BaBar collaboration has reported a search on $B^+ \to Y(4260)(\to J/\psi \pi^+ \pi^-) K^+$~\cite{ref:PRD73011101}. They found $128\pm42$ signal events using a data sample of 211 fb$^{-1}$, with a statistical significance of 3.1 standard deviations ($\sigma$).
They set an upper limit on $\mathcal{B}(B^+ \to Y(4260) K^+) \times \mathcal{B}(Y(4260) \to J/\psi \pi^+ \pi^-) < 2.9 \times 10^{-5}$ at 95\% confidence level (C.L.).

In our analysis, both charged and neutral $B$'s are considered. The $Y(4260)$ candidates are reconstructed from $J/\psi \pi^+ \pi^-$, where $J/\psi \to \ell^+ \ell^- (\ell \in \{e,\mu \})$. $B \to \psi(2S) K$ and $B \to X(3872) K$ are used as control samples to validate and calibrate our Monte Carlo (MC) results since they have similar topology and larger statistics to our mode.
An unbinned extended maximum likelihood (UML) fit is performed to the distribution of the energy difference $\Delta E = \sum_i E_i - E_\text{beam}$, where $E_\text{beam}$ and $E_i$ are the beam energy and the energy of the $i^{th}$ daughter particle in the center-of-mass frame, respectively. The statistical weight for each candidate to be a signal decay is determined using the $_sPlot$ technique~\cite{ref:splot}, and the signal yield for the intended resonance is extracted from an UML fit to the weighted $M_{J/\psi \pi^+ \pi^-}$ distribution.

The observed signal yields for the charged and neutral $B\to Y(4260)(\to J/\psi \pi^+ \pi^-) K$ decays are $179\pm53^{+55}_{-41}$ and $39\pm28^{+7}_{-31}$, respectively; the signal significances are $2.1\sigma$ and $0.9\sigma$, respectively.
Since there are no any significant signals, we set the upper limit (U.L.) for the branching fraction product to be $1.4\times10^{-5}$ and $1.7\times10^{-5}$ at the  90\% C.L., and $1.56\times10^{-5}$ and $2.16\times10^{-5}$ at the 95\% C.L. for the charged and neutral decays, respectively. The results are summarized in Table.~\ref{table:Y4260}.
They are the most stringent upper limits to date. The upper limit for the charged decay is consistent with the BaBar's previous result, and the upper limit for the neutral decay is given for the first time.

\begin{table}[htbp]
\caption{\label{table:Y4260}Results for the $B \to Y(4260) K$ search. $\epsilon$ represents the selection efficiency, and $N_\text{sig}$ represents the signal yields.}
\begin{ruledtabular}
\begin{tabular}[t]{cccccc}
Decay      & $\epsilon (\%)$  &$N_\text{sig}$ & Significance & U.L. (90\% C.L.) & U.L. (95\% C.L.)\\
\hline
$B^+\to Y(4260) K^+$, $Y(4260)\to J/\psi \pi^+ \pi^-$ & 19.8 & $179\pm53^{+55}_{-41}$ & $2.1\sigma$ & $1.4\times10^{-5}$ & $1.56\times10^{-5}$ \\
$B^0\to Y(4260) K^0$, $Y(4260)\to J/\psi \pi^+ \pi^-$ & 10.6 & $39\pm28^{+7}_{-31}$ & $0.9\sigma$ & $1.7\times10^{-5}$ & $2.16\times10^{-5}$ \\
\end{tabular}
\end{ruledtabular}
\end{table}

\section{Evidence for $B \to h_c K$ and observation of $\eta_c(2S) \to p\bar{p}\pi^+\pi^-$\cite{ref:hcK}}
The decays $B^+ \to h_c K^+$, $B^+ \to \chi_{c0} K^+$ and $B^+ \to \chi_{c2} K^+$ are all suppressed by factorization~\cite{ref:ZPC34103,ref:PRD66037503}. However, the current world average $\mathcal{B}(B^+ \to \chi_{c0} K^+) = (1.49^{+0.15}_{-0.14})\times 10^{-4}$~\cite{ref:PDG2018} is 
not strongly suppressed and only slightly smaller than the factorization-allowed process $\mathcal{B}(B^+ \to \chi_{c1} K^+) = (4.84\pm0.23)\times 10^{-4}$. Before the first experimental searches, it was expected that $\mathcal{B}(B^+ \to h_c K^+) \approx \mathcal{B}(B^+ \to \chi_{c0} K^+)$~\cite{ref:PRD66037503}. Before our search, the best upper limit was $\mathcal{B}(B^+ \to h_c K^+) < 3.8\times 10^{-5}$ at 90\% C.L., which was obtained in the $h_c$ search by Belle in 2006~\cite{ref:PRD74012007}. Although the LHCb collaboration also set the upper limit on the branching fraction product $\mathcal{B}(B^+ \to h_c K^+)\times \mathcal{B}(h_c \to p \bar{p}) < 6.4\times 10^{-8}$ at 95\% C.L.~\cite{ref:EPJC732462}, it does not give a stronger constraint since the decay $h_c \to p\bar{p}$ has never been observed and $\mathcal{B}(h_c \to p \bar{p}) < 1.5\times 10^{-4}$ at 90\% C.L.~\cite{ref:PDG2018}.
After the experimental upper limit of $\mathcal{B}(B^+ \to h_c K^+)$ was set, some new theoretical predictions based on different approaches were made. In the QCD factorization approach, the branching fraction is calculated to be $2.7\times10^{-5}$~\cite{ref:hepph0607221}, while using perturbative QCD (pQCD) approach the result is $\mathcal{B}(B^+ \to h_c K^+) = 3.6\times10^{-5}$~\cite{ref:PRD74114029}. Both values are slightly below the previous experimental upper limit.

In our analysis, both charged and neutral $B$'s are considered. The $h_c$ candidates are reconstructed from $\eta_c \gamma$ and $p\bar{p}\pi^+\pi^-$, where the latter decay channel was recently observed by BESIII~\cite{ref:PRD99.072008}. The $\eta_c$ candidates are reconstructed in ten different decay channels: $K^+K_S^0\pi^-$, $K^+K^-\pi^0$, $K_S^0K_S^0\pi^0$, $K^+K^-\eta$, $K^+K^-K^+K^-$, $\eta'(\to \eta \pi^+ \pi^-$) $\pi^+ \pi^-$, $p\bar{p}$, $p\bar{p}\pi^0$, $p\bar{p}\pi^+\pi^-$, and $\Lambda\bar{\Lambda}$. 
Also, the integrated luminosity is 2.8 times greater than the luminosity used previously~\cite{ref:PRD74012007}.
Multivariate analysis (MVA) with a multilayer perceptron neural network~\cite{ref:MLP} is used for each channel to separate the signal events from the background events. A simultaneous UML fit to the invariant mass of the $h_c$ candidate is performed to $h_c \to \eta_c \gamma$ signal,  $h_c \to p\bar{p}\pi^+\pi^-$ background, and $h_c \to p\bar{p}\pi^+\pi^-$ signal distributions.
The decays of other charmonium states to $p\bar{p}\pi^+\pi^-$ channel are also studied.

Evidence for the charged decay $B^+ \to h_c K^+$ is found with a significance of $4.8\sigma$, and its branching fraction is measured to be $(3.7^{+1.0 +0.8}_{-0.9-0.8}) \times 10^{-5}$. It is consistent with and supersedes the existing upper limit, and  it also agrees with the theoretical predictions.
The neutral decay $B^0 \to h_c K^0_S$ is not found, and we set the upper limit for its branching fraction to be $1.4\times10^{-5}$ at 90\% C.L. The upper limit for the neutral decay is given for the first time.
Furthermore, we observed the decay $\eta_c(2S) \to p\bar{p}\pi^+\pi^-$ for the first time with $12.1\sigma$ significance, by studying the $p\bar{p}\pi^+\pi^-$ invariant mass distribution in $B^+ \to (p\bar{p}\pi^+\pi^-) K^+$ channel. Other charmonium signals are consistent with the current world-average values. The results are summarized in Table.~\ref{table:hcK}.

\begin{table}[htbp]
\caption{\label{table:hcK}Results for $B \to (c\bar{c}) K$ measurements. The values or confidence intervals are at 90\% C.L. }
\begin{ruledtabular}
\begin{tabular}[t]{cccc}
Branching fraction                              & Significance & Value or confidence interval       & World-average value~\cite{ref:PDG2018}\\
\hline
$\mathcal{B}(B^+ \to h_c K^+)$ & $4.8\sigma$ & $(3.7^{+1.0 +0.8}_{-0.9-0.8}) \times 10^{-5}$ & $<3.8\times10^{-5}$ \\
$\mathcal{B}(B^+ \to \eta_c K^+)\times \mathcal{B}(\eta_c \to p\bar{p}\pi^+\pi^-)$ & $20.1\sigma$ & $(39.4^{+4.1 +2.2}_{-3.9-1.8}) \times 10^{-7}$ & $(57.8\pm20.2)\times10^{-7}$ \\
$\mathcal{B}(B^+ \to J/\psi K^+)\times \mathcal{B}(J/\psi \to p\bar{p}\pi^+\pi^-)$ & $33.9\sigma$ & $(56.4^{+3.3 +2.7}_{-3.2-2.5}) \times 10^{-7}$ & $(60.6\pm5.3)\times10^{-7}$ \\
$\mathcal{B}(B^+ \to \chi_{c0} K^+)\times \mathcal{B}(\chi_{c0} \to p\bar{p}\pi^+\pi^-)$ & $6.0\sigma$ & $(3.7^{+1.2 +0.2}_{-1.0-0.3}) \times 10^{-7}$ & $(3.1\pm1.1)\times10^{-7}$ \\
$\mathcal{B}(B^+ \to \chi_{c1} K^+)\times \mathcal{B}(\chi_{c1} \to p\bar{p}\pi^+\pi^-)$ & $4.9\sigma$ & $(4.7^{+1.3 +0.4}_{-1.2-0.2}) \times 10^{-7}$ & $(2.4\pm0.9)\times10^{-7}$ \\
$\mathcal{B}(B^+ \to \chi_{c2} K^+)\times \mathcal{B}(\chi_{c2} \to p\bar{p}\pi^+\pi^-)$ & $0.3\sigma$ & $<1.9 \times 10^{-7}$ & $(0.15\pm0.06)\times10^{-7}$ \\
$\mathcal{B}(B^+ \to \eta_c(2S) K^+)\times \mathcal{B}(\eta_c(2S) \to p\bar{p}\pi^+\pi^-)$ & $12.3\sigma$ & $(11.2^{+1.8 +0.5}_{-1.6-0.7}) \times 10^{-7}$ & not seen \\
$\mathcal{B}(B^+ \to \psi(2S) K^+)\times \mathcal{B}(\psi(2S) \to p\bar{p}\pi^+\pi^-)$ & $5.0\sigma$ & $[0.5,3.5] \times 10^{-7}$ & $(3.7\pm0.3)\times10^{-7}$ \\
\hline
$\mathcal{B}(B^0 \to h_c K^0_S)$ & $0.7\sigma$ & $<1.4 \times 10^{-5}$ & not seen \\
$\mathcal{B}(B^0 \to \eta_c K^0_S)\times \mathcal{B}(\eta_c \to p\bar{p}\pi^+\pi^-)$ & $12.5\sigma$ & $(19.0^{+3.2 +1.3}_{-2.9-4.7}) \times 10^{-7}$ & $(20.9\pm7.8)\times10^{-7}$ \\
$\mathcal{B}(B^0 \to J/\psi K^0_S)\times \mathcal{B}(J/\psi \to p\bar{p}\pi^+\pi^-)$ & $20.8\sigma$ & $(24.3^{+2.3 +1.2}_{-2.2-1.3}) \times 10^{-7}$ & $(26.2\pm2.4)\times10^{-7}$ \\
$\mathcal{B}(B^0 \to \chi_{c0} K^0_S)\times \mathcal{B}(\chi_{c0} \to p\bar{p}\pi^+\pi^-)$ & $0.6\sigma$ & $<1.3 \times 10^{-7}$ & $(1.5\pm0.6)\times10^{-7}$ \\
$\mathcal{B}(B^0 \to \chi_{c1} K^0_S)\times \mathcal{B}(\chi_{c1} \to p\bar{p}\pi^+\pi^-)$ & $4.5\sigma$ & $(3.7^{+1.2 +0.3}_{-1.0-0.2}) \times 10^{-7}$ & $(1.0\pm0.4)\times10^{-7}$ \\
$\mathcal{B}(B^0 \to \chi_{c2} K^0_S)\times \mathcal{B}(\chi_{c2} \to p\bar{p}\pi^+\pi^-)$ & $2.5\sigma$ & $[0.7,3.8]\times 10^{-7}$ & not seen \\
$\mathcal{B}(B^0 \to \eta_c(2S) K^0_S)\times \mathcal{B}(\eta_c(2S) \to p\bar{p}\pi^+\pi^-)$ & $5.9\sigma$ & $(4.2^{+1.4 +0.3}_{-1.2-0.3}) \times 10^{-7}$ & not seen \\
$\mathcal{B}(B^0 \to \psi(2S) K^0_S)\times \mathcal{B}(\psi(2S) \to p\bar{p}\pi^+\pi^-)$ & $2.8\sigma$ & $<1.9 \times 10^{-7}$ & $(1.7\pm0.2)\times10^{-7}$ \\
\end{tabular}
\end{ruledtabular}
\end{table}

\section{Search for $X(3872/3915) \to \chi_{c1} \pi^0$\cite{ref:Xtochic1pi0}}
The decay $X(3872) \to \chi_{c1} \pi^0$ was recently observed by BESIII in $e^+e^- \to \chi_{c1}\pi^0\gamma$~\cite{ref:PRL.122.202001}. 
If the charmonium $\chi_{c1}(2P)$ component dominates the $X(3872)$ structure, the pionic transition $X(3872) \to \chi_{c1} \pi^0$ is expected to be very small due to isospin breaking by the light quark mass~\cite{ref:PRD77.014013}.
However, according to BESIII's result, $\frac{\mathcal{B}(X(3872)\to  \chi_{c1} \pi^0)}{\mathcal{B}(X(3872)\to J/\psi \pi^+ \pi^- ) }= 0.88^{+0.33}_{-0.27} \pm 0.10$ is large compared to $\frac{\mathcal{B}(\psi(2S)\to J/\psi \pi^0) }{ \mathcal{B}(\psi(2S) \to J/\psi \pi^+ \pi^-) }= 3.66\times10^{-3}$, which disfavors the $\chi_{c1}(2P)$ interpretation. 
The $X(3915)$ state was first observed by Belle in $B \to X(3915)K \to J/\psi \omega K$~\cite{ref:PRL94.182002}.
For a pure charmonium $\chi_{c0}(2P)$ scenario, its width is too narrow (expect > 100 MeV/$c^2$~\cite{ref:PRD86.091501}, measured (20$\pm$5) MeV/$c^2$~\cite{ref:PDG2018}) and its branching fraction is too large (should be suppressed by OZI rule~\cite{ref:PRD91.057501}). If $X(3915)$ is a non-conventional state, single pion transitions may be enhanced.

In our analysis, the $X(3872)$ and $X(3915)$ candidates are produced by the $B^+ \to X(3872/3915) K^+$ decay. The $\chi_{c1}$ candidates are reconstructed from $J/\psi \gamma$, where $J/\psi \to \ell^+ \ell^- (\ell \in \{e,\mu\})$. The signal yield is extracted from an UML fit to the $\Delta E$ distribution, 
and the $X(3872/3915)$ signal yield is extracted from an UML fit to the weighted $M_{\chi_{c1}\pi^0}$ distribution produced by the $_sPlot$ technique. Events from $B^+ \to \chi_{c1} K^{*+}$ are vetoed by rejecting events with 791.8 MeV/$c^2 < M_{K^+\pi^0} < $ 991.8 MeV/$c^2$.

The observed signal yields for the $X(3872)$ and $X(3915)$ modes are $2.7\pm5.5$ and $42\pm14$, respectively; the signal significances are $0.3\sigma$ and $2.3\sigma$, respectively.
Since there are no any significant signals, we set the upper limit (U.L.) for the branching fraction product $\mathcal{B}(B^+ \to X(3872) K^+) \times \mathcal{B}(X(3872) \to \chi_{c1} \pi^0) < 8.1\times10^{-6}$ and $\mathcal{B}(B^+ \to X(3915) K^+) \times \mathcal{B}(X(3915) \to \chi_{c1} \pi^0)<3.8\times10^{-5}$ at the 90\% C.L.,
which are compatible with the $D^0D^{*0}$ + $\chi_{c1}(2P)$ admixture scenario for $X(3872)$~\cite{ref:PRD77.014013}. 
We also obtain that $\frac{\mathcal{B}(X(3872) \to \chi_{c1} \pi^0) }{\mathcal{B}(X(3872) \to J/\psi \pi^+\pi^-) }< 0.97$ at the 90\% C.L.,
which does not contradict the BESIII result.
The results are summarized in Table.~\ref{table:X3915}, and these information can be used to constrain the molecular/tetraquark component of the $X$ states.

\begin{table}[htbp]
\caption{\label{table:X3915}Results for the $X(3872/3915) \to \chi_{c1} \pi^0$ search. $\epsilon$ represents the selection efficiency, and $N_\text{sig}$ are the signal yields.}

\begin{ruledtabular}
\begin{tabular}[t]{ccccc}
Decay      & $\epsilon (\%)$  &$N_\text{sig}$ & Significance & U.L. (90\% C.L.)\\
\hline
$B^+ \to X(3872) K^+$, $X(3872) \to \chi_{c1} \pi^0 $ & 5.35 & $2.7\pm5.5$ & $0.3\sigma$ & $8.1\times10^{-6}$  \\
$B^+ \to X(3915) K^+$, $X(3915) \to \chi_{c1} \pi^0 $ & 5.37 & $42\pm14$ & $2.3\sigma$& $3.8\times10^{-5}$ \\
\end{tabular}
\end{ruledtabular}
\end{table}

\section{Search for $B^0 \to X(3872) \gamma$\cite{ref:X3872gamma}}
Predictions of $B^0 \to (c\bar{c}) \gamma$ branching fractions depend on the factorization approach of QCD interactions. For example, the branching fraction of $B^0 \to J/\psi \gamma$ is predicted to be $7.65\times10^{-9}$ and $4.5\times10^{-7}$ using QCD factorization~\cite{ref:EPJC34.291} and pQCD approach~\cite{ref:PRD74.097502}, respectively. The current upper limit for $B^0 \to J/\psi \gamma$ is $1.5 \times 10^{-6}$ at 90\% confidence level~\cite{ref:PDG2018}. Possible new physics enhancements of the branching fractions for such decay may be due to right-handed currents~\cite{ref:EPJC34.291} or nonspectator intrinsic charm in $B^0$~\cite{ref:PRD65.054016}.
Since $X(3872)$ may not be a pure charmonium state, the branching fraction of $B^0 \to X(3872) \gamma$ should be smaller than that of $B^0 \to J/\psi \gamma$. This is the first search of the decay $B^0 \to X(3872) \gamma$.

In our analysis, the $X(3872)$ candidates are reconstructed from $J/\psi \pi^+ \pi^-$, where $J/\psi \to \ell^+ \ell^- (\ell \in \{e,\mu \})$.
MVA with a neural network~\cite{ref:NeuroBayes} is used to separate the signal events from the background events. 
$B^0 \to K^0_S \pi^+ \pi^- \gamma$, $B^0 \to J/\psi K^0_S$, and $B^0 \to \psi(2S) K^0_S$ are used as control samples to validate and calibrate our MC results. 
$B$ meson candidates are identified with the energy difference $\Delta E$ and a modified beam-energy-constrained mass $M_\text{bc} = \sqrt{\Big(\frac{E^*_\text{beam}}{c^2}\Big)^2 - \Big( \frac{\vec{P}^*_{X}}{c}+\frac{\vec{P}^*_\gamma}{|\vec{P}^*_\gamma| c^2}(E^*_\text{beam}-E^*_\text{X})\Big)^2}$, 
in which $\vec{P}^*_{X}$ and $E^*_{X}$ are the reconstructed momentum and energy of the $X(3872)$ candidate, and $\vec{P}^*_\gamma$ is the reconstructed momentum of the photon candidate.
A signal region is defined by ranging $\Delta E$ and $M_\text{bc}$, and the Feldman-Cousins counting method~\cite{ref:PRD57.3873,ref:CPC181.683} is used to obtain the upper limit.

The observed number of events $N_\text{evt}$ in the signal region are both 9 for dimuon and dielectron channels, respectively. Such numbers are about the same as the expected background numbers $N_\text{bkg}$ in the same region, which are 9.3 and 12.1 for dimuon and dielectron channels, respectively.
Since there are no any significant signals, we set the upper limit for the branching fraction product 
$\mathcal{B}(B^0 \to X(3872) \gamma)\times \mathcal{B}(X(3872) \to J/\psi \pi^+ \pi^-) < 5.1\times10^{-7}$ at the 90\% C.L.
The results are summarized in Table.~\ref{table:X3872gamma}. This upper limit  is given for the first time.

\begin{table}[htbp]
\caption{\label{table:X3872gamma}Results for the $B^0 \to X(3872) \gamma$ search.}
\begin{ruledtabular}
\begin{tabular}[t]{lcccc}
Channel & Observed $N_\text{evt}$   & Expected $N_\text{bkg}$ & Efficiency (\%)& 90\% U.L.\\
\hline
Di\-muon & 9 & 9.3 & $16.8\pm0.01$ & $9.2\times10^{-7}$ \\
Di\-electron & 9 & 12.1 & $14.5\pm0.01$ & $6.8\times10^{-7}$ \\
Total & 18 & 21.4 & -- & $5.1\times10^{-7}$ \\
\end{tabular} 
\end{ruledtabular}
\end{table}

\section{Observation of $e^+e^- \to \gamma \chi_{c1}$ and search for $e^+e^- \to \gamma \chi_{c0}$, $\gamma \chi_{c2}$ and $\gamma \eta_c$\cite{ref:gammachic1}}
Electromagnetic quarkonium production serves as a good testing ground for nonrelativistic quantum chromodynamics (NRQCD) predictions for its relative simplicity.
The BESIII experiments measured the cross sections of $e^+e^- \to \gamma \chi_{cJ} (J=0,1,2)$ at $\sqrt{s} = 4.01$, 4.23, 4.26, and 4.36 GeV, as well as the cross sections of $e^+e^- \to \gamma \eta_c$ at the same energies and at 4.42 and 4.60 GeV additionally~\cite{ref:CPC39.041001,ref:PRD96.051101}.
However, at none of those individual energy points do the statistical significances for $\chi_{cJ}$ or $\eta_c$ production exceed $3\sigma$. When the data from all energy points are combined, the statistical significances for $\chi_{c1}$, $\chi_{c2}$, and $\eta_c$ production are $3.0\sigma$, $3.4\sigma$, and $>3.6\sigma$, respectively. Furthermore, BESIII has reported evidence for $e^+e^- \to X(3872) \gamma$~\cite{ref:PRL112.092001}, and a precise measurement of $e^+e^- \to \gamma \chi_{cJ}$ and $\gamma \eta_c$ will be useful to understand the $C$-even quarkonia and the exotic XYZ states including $X(3872)$~\cite{ref:PRD90.034020,ref:arxiv1310.8597,ref:PRD79.094504}.

In our analysis, the $\chi_{cJ}$ candidates are reconstructed from $J/\psi \gamma$, where $J/\psi \to \mu^+ \mu^-$. The $\eta_c$ candidates are reconstructed from $K^+K_S^0\pi^-$, $\pi^+\pi^-K^+K^-$, $\pi^+\pi^-\pi^+\pi^-$, $K^+K^-K^+K^-$, and $3(\pi^+\pi^-)$. The analysis is performed on $\sqrt{s} =$ 10.52, 10.68, 10.867 GeV, and MVA with a neural network~\cite{ref:NeuroBayes} is also used to suppress the background events.
Corrections due to initial-state radiation (ISR) are taken into account by assuming $\sigma(e^+e^- \to \gamma \chi_{cJ}/\eta_c) \sim 1/s^n$, where $n=$ 1.4, 2.1, 2.4, and 1.3 for $\chi_{c0}$, $\chi_{c1}$, $\chi_{c2}$, and $\eta_c$, respectively~\cite{ref:PRD97.094504,ref:JHEP01.2018.091}.
The signal yields of the $\chi_{cJ}$ and $\eta_c$ candidates are extracted from an UML fit to their invariant mass distributions, where the five $\eta_c$ final states are performed by a simultaneous fit. 

We observed a clear signal of $\chi_{c1}$ signal at $\sqrt{s} = 10.58$ GeV with a significance of $5.1\sigma$ including systematic uncertainties, and the Born cross section is measured to be $(17.3^{+4.2}_{-3.9} ({\rm stat.}) \pm 1.7 ({\rm syst.}))$ fb. For other data samples, the signal is not evident and their corresponding upper limits are given at 90\% C.L.
The results are summarized in Table~\ref{table:gammachic1}. 
Together with the BESIII measurements~\cite{ref:CPC39.041001} at lower center-of-mass energies, the $s$-dependency of the Born cross section for $e^+e^- \to \gamma \chi_{c1}$ is obtained to be $1/s^{2.1^{+0.3}_{-0.4}\pm0.3}$.

\begin{table}[htbp]
\caption{\label{table:gammachic1}Results for the $e^+e^- \to \gamma \chi_{cJ}$ and $e^+e^- \to \gamma \eta_c$ search. $N_\text{sig}$ ($N^\text{UL}_\text{sig}$) represents the (upper limit of) signal yields, $\sigma$ is the signal significance, $\epsilon$ represents the selection efficiency, and $\sigma_B$ ($\sigma^\text{UL}_B$) represents the (upper limit of) Born cross sections. }
\begin{ruledtabular}
\begin{tabular}[t]{lccccccc}
Channel & $\sqrt{s}$ (GeV) & $N_\text{sig}$ & $N^\text{UL}_\text{sig}$ (90\% C.L.) & $\sigma$ & $\epsilon$(\%) &$\sigma_B$ (fb) & $\sigma^\text{UL}_B$ (90\% C.L.) \\
\hline
$e^+e^- \to \gamma \chi_{c0}$ & 10.52 & $2.9^{+4.0}_{-3.3}$      &9.6    &0.9 &19.0 & $286.2^{+394.7}_{-325.6}\pm30.7$ & 957.2 \\
$e^+e^- \to \gamma \chi_{c1}$ &           & $4.8^{+3.6}_{-2.9}$      &10.4  &1.9 &20.8 & $16.2^{+12.1}_{-9.8}\pm1.4$ & 34.9 \\
$e^+e^- \to \gamma \chi_{c2}$ &           & $-0.8^{+2.3}_{-1.6}$     &4.5    &--   &19.9 & $-5.0^{+14.3}_{-10.0}\pm0.6$ & 28.9 \\
$e^+e^- \to \gamma \eta_c$     &           & $6.8^{+14.8}_{-14.3}$  &30.8  &0.5 &0.79 & $9.0^{+19.5}_{-18.8}\pm1.0$ & 40.6 \\
$e^+e^- \to \gamma \chi_{c0}$ & 10.58 & $-1.6^{+9.8}_{-8.9}$     &16.5  &--   &18.9 & $-20.0^{+122.3}_{-111.0}\pm2.6$ & 205.9 \\
$e^+e^- \to \gamma \chi_{c1}$ &           & $39.0^{+9.5}_{-8.8}$    &--       &5.2 &19.9 & $17.3^{+4.2}_{-3.9}\pm1.7$ & -- \\
$e^+e^- \to \gamma \chi_{c2}$ &           & $-8.7^{+5.7}_{-5.0}$     &7.2    &--   &19.8 & $-6.8^{+4.5}_{-3.9}\pm1.4$ & 5.7 \\
$e^+e^- \to \gamma \eta_c$     &           & $67.2^{+42.0}_{-39.2}$&125.9&1.8 &0.78 & $11.3^{+7.0}_{-6.6}\pm1.5$ & 21.1 \\
$e^+e^- \to \gamma \chi_{c0}$ &10.867& $-1.3^{+4.0}_{-3.2}$     &7.0    &--   &17.7 & $-101.4^{+312.0}_{-249.6}\pm9.5$ & 543.7 \\
$e^+e^- \to \gamma \chi_{c1}$ &           & $1.9^{+3.4}_{-2.6}$      &7.9    &0.7 &16.8 & $5.8^{+10.5}_{-8.0}\pm0.8$ & 24.3 \\
$e^+e^- \to \gamma \chi_{c2}$ &           & $-2.8^{+3.2}_{-2.4}$     &5.3    &--   &16.3 & $-15.7^{+17.9}_{-13.4}\pm2.3$ & 30.3 \\
$e^+e^- \to \gamma \eta_c$     &           & $12.3^{+18.2}_{-17.4}$&42.3  &0.9 &0.76 & $12.3^{+17.3}_{-18.1}\pm1.1$ & 42.2 \\
\end{tabular} 
\end{ruledtabular}
\end{table}

\section{Measurement of the $e^+e^- \to \Upsilon(nS) \pi^+ \pi^-$ Cross Sections\cite{ref:Upsilonpipi}}
Above the $B\bar{B}$ threshold, the vector bottomonium states $\Upsilon(4S)$, $\Upsilon(10860)$ and $\Upsilon(11020)$ have properties that are unexpected for pure $b\bar{b}$ bound states~\cite{ref:MPLA32.1750025}. 
Compared to ordinary bottomonium states, their transition to lower bottomonia with light hadron emission have much higher rates, and some of them even strongly violate the Heavy Quark Spin Symmetry.
Possible explanations for these unexpected properties including contribution of hadron loops (equivalently, presence of a $B^{(*)}_{(s)}\bar{B}^{(*)}_{(s)}$ admixture)~\cite{ref:PRD77.074003,ref:PLB671.55,ref:PRD85.034024} or presence of other exotic states (e.g. compact tetraquarks~\cite{ref:PRL104.162001} or hadrobottomonia~\cite{ref:PLB666.344}).
Besides, $\Upsilon(3, 4D)$ states are predicted in the region of the $\Upsilon(4S)$, $\Upsilon(10860)$ and $\Upsilon(11020)$ levels~\cite{ref:EPJC71.1825,ref:PRD92.054034}.  
Although the electron widths of the $D$-wave states are expected to be quite small for bottomonia below the $B\bar{B}$ threshold~\cite{ref:PRD28.1132}, 
they can be significantly enhanced above the open-flavor thresholds due to $B$-meson loops~\cite{ref:PAN73.138}.
Furthermore, recent study of $e^+e^-  \to  \Upsilon(nS) \pi^+ \pi^-$ in Belle show a small hint of new structure at $\sqrt{s} = 10.77$ GeV~\cite{ref:PRD93.011101}. 
It is of interest to study more channels and to improve the accuracy of the previously measured cross sections.

In our analysis, the $\Upsilon(nS) (n=1,2,3)$ candidates are reconstructed from  $\ell^+ \ell^- (\ell \in \{ \mu, e\})$.
We scanned over the data with about 1 fb$^{-1}$ per point collected in the  energy range from 10.63 GeV to 11.02 GeV.
The $\Upsilon(10860)$ on-resonance data samples with a total luminosity of 121 fb$^{-1}$ are also used, 
and they were collected in five different periods with slightly different c.m. energies between 10.864 GeV and 10.868 GeV.
Finally, we use the continuum data sample collected at 10.52 GeV with a total luminosity of 60 fb$^{-1}$. 
Signal yields are extracted via fitting to the $M_\text{recoil}(\pi^+\pi^-)$ distribution, instead of the counting method used in the previous study~\cite{ref:PRD93.011101}.
Additional information about the cross section shapes are obtained by using the ISR process with the high statistics $\Upsilon(10860)$ on-resonance data. 
The fully reconstructed events are selected with an energy balance requirement $|M_\text{recoil}(\pi^+\pi^-)-M(\ell^+\ell^-)| < 150$ MeV.

A new measurement of the energy dependence of the cross sections for $e^+e^- \to \Upsilon(nS) \pi^+ \pi^-$ which supersedes the previous Belle result~\cite{ref:PRD93.011101} is reported.
Furthermore, we observed a new structure in the energy dependence, with a global significance of 6.7$\sigma$ including the systematic uncertainties. 
Such new structure may be explained by resonances of the not-yet-observed $\Upsilon(3D)$ (enhancement of $S$-$D$ mixing)~\cite{ref:PAN73.138} or an exotic state (e.g. a compact tetraquark~\cite{ref:PRL104.162001} or hadro\-bottomonium~\cite{ref:PLB666.344}).
It may also be a non-resonant effect due to some complicated rescattering.
Measurements of $\Upsilon(10860)$ and $\Upsilon(11020)$ parameters with improved accuracy are also reported.
In the continuum data sample at $\sqrt{s}=10.52$ GeV, a clear signal for the $e^+e^- \to \Upsilon(1S) \pi^+ \pi^-$  process is evident. Its significance including systematic uncertainties is larger than 3.5$\sigma$, and the corresponding cross section is determined to be $42^{+17}_{-15}$ fb. 

\begin{table}[htbp]
\caption{\label{table:Ypipi}Results for the $e^+e^- \to \Upsilon(nS) \pi^+ \pi^-$ cross-section measurement. $M$ and $\Sigma$ represent the measured masses and widths, where the first and second uncertainties are statistical and systematic, respectively. Ranges of the $\Sigma_{ee}\times\mathcal{B}$ values are shown from the lowest to the highest solution.}
\begin{ruledtabular}
\begin{tabular}[t]{lccccc}
 & $M$   &   $\Sigma$ & \multicolumn{3}{c}{$\Sigma_{ee}\times\mathcal{B}$ (eV)}\\
 & (MeV/$c^2$)   &(MeV) & $\Upsilon(1S) \pi^+ \pi^-$ & $\Upsilon(2S) \pi^+ \pi^-$ & $\Upsilon(3S) \pi^+ \pi^-$\\
\hline
$\Upsilon(10860)$ & $10885.3\pm1.5^{+2.2}_{-0.9}$ & $36.6^{+4.5+0.5}_{-3.9-1.1}$ & 0.75 -- 1.43 & 1.35 -- 3.80 & 0.43 -- 1.03\\
$\Upsilon(11020)$ & $11000.0^{+4.0+1.0}_{-4.5-1.3}$ & $23.8^{+8.0+0.7}_{-6.8-1.8}$ & 0.38 -- 0.54 & 0.13 -- 1.16 & 0.17 -- 0.49\\
New structure        & $10752.7\pm5.9^{+0.7}_{-1.1}$ & $35.5^{+17.6+3.9}_{-11.3-3.3}$ & 0.12 -- 0.47 & 0.53 -- 1.22 & 0.21 -- 0.26\\
\end{tabular} 
\end{ruledtabular}
\end{table}

\section{Conclusion}
We review some results on the study of charmonia and bottomonia-like particles at Belle, and the main results are summarized below. 
(1) the first observation of the decay $\eta_c(2S) \to p \bar{p} \pi^+ \pi^-$, the process 
$e^+e^- \to \gamma \chi_{c1}$ at $\sqrt{s}=10.58$ GeV, 
and a new structure in the energy dependence of $e^+e^- \to \Upsilon(nS) \pi^+ \pi^-$ process;
(2) the evidence of the process $e^+e^- \to \Upsilon(1S) \pi^+ \pi^-$ at $\sqrt{s}=10.52$ GeV and of the decay $B^+ \to h_c K^+$;
(3) the energy dependence of the cross sections for $e^+e^- \to \gamma \chi_{c1}$ and $e^+e^- \to \Upsilon(nS) \pi^+ \pi^-$;
(4) the 90\% C.L. upper limits are set for the search for some processes related to charmonium and charmonium-like states such as $B \to Y(4260) K$, $B^0 \to h_c K^0_s$, $X(3872/3915) \to \chi_{c1} \pi^0$, $B^0 \to X(3872) \gamma$, $e^+e^- \to \gamma \chi_{cJ}$
, and $e^+e^- \to \gamma \eta_c$.
We expect 40 times of integrated luminosity in Belle II, and the measurements for these processes can be further improved with higher precision.

\begin{acknowledgments}
We thank the KEKB group for excellent operation of the
accelerator; the KEK cryogenics group for efficient solenoid
operations; and the KEK computer group, the NII, and 
PNNL/EMSL for valuable computing and SINET5 network support.  
We acknowledge support from MEXT, JSPS and Nagoya's TLPRC (Japan);
ARC (Australia); FWF (Austria); NSFC and CCEPP (China); 
MSMT (Czechia); CZF, DFG, EXC153, and VS (Germany);
DST (India); INFN (Italy); 
MOE, MSIP, NRF, RSRI, FLRFAS project and GSDC of KISTI and KREONET/GLORIAD (Korea);
MNiSW and NCN (Poland); MSHE (Russia); ARRS (Slovenia);
IKERBASQUE (Spain); 
SNSF (Switzerland); MOE and MOST (Taiwan); and DOE and NSF (USA).
\end{acknowledgments}

\nocite{*}
\bibliography{ms}

\end{document}